%% file: paper-main.tex
\documentclass{sig-alternate-2013}

\usepackage{amsmath}
\usepackage{xcolor}
\usepackage{graphicx}
\usepackage{subfigure}
\graphicspath{{Images/}}
\usepackage{float}
\usepackage{url}
\usepackage{latexsym}
\usepackage{color}
\usepackage{pgf}	
\usepackage{multicol}
\usepackage[labelfont=bf,font=bf]{caption}
\usepackage{lipsum}
\usepackage{enumerate}
\usepackage{xspace}
\usepackage{booktabs}

\usepackage{hyperref}
\usepackage[hyphenbreaks]{breakurl}

\permission{\copyright 2017 International World Wide Web Conference Committee \\ (IW3C2), published under Creative Commons CC BY 4.0 License.}
\conferenceinfo{WWW 2017,}{April 3--7, 2017, Perth, Australia.}
\copyrightetc{ACM \the\acmcopyr}
\crdata{978-1-4503-4913-0/17/04. \\
http://dx.doi.org/10.1145/3038912.3052652 \\
\includegraphics{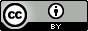}
}
\numberofauthors{4}
\author{
\alignauthor
\hspace{-0.2cm}Rahmtin Rotabi\\
       \hspace{-0.2cm}\affaddr{Cornell University}\\
       \email{\hspace{-0.2cm}rahmtin@cs.cornell.edu}
\alignauthor
\mbox{\hspace{-0.7cm}Cristian Danescu-Niculescu-Mizil}
       \affaddr{Cornell University}\\
       \email{cristian@cs.cornell.edu}
\alignauthor
\hspace{0.1cm}Jon Kleinberg\\
  \hspace{0.1cm}     \affaddr{Cornell University}\\
   \mbox{\hspace{0.3cm}\email{kleinber@cs.cornell.edu}}
}	
\newcommand{\cut}[1]{}

\input{macros.tex}

\title{Competition and Selection Among Conventions}

\begin{document}

\pgfmathsetseed{\number\pdfrandomseed}
\pgfmathparse{rnd}

\maketitle
\begin{abstract}
\input{000abstract}
\end{abstract}

\keywords{conventions, innovations, diffusion of information, arXiv}

\section{Introduction}
\label{sec:intro}
\input{010intro}
\section{Further Related Work}
\label{sec:related}
\input{020related}

\section{Data}
\label{sec:data}
\input{030data}
\input{040method}

\section{Discussion}
\label{sec:discussion}
\input{060discussion}

\input{070acknowledgement}

\bibliographystyle{abbrv}
\bibliography{refs}
\end{document}

%% file: macros.tex
\newcommand{\name}{\texttt{name}\xspace}
\newcommand{\names}{\texttt{names}\xspace}

\newcommand{\Names}{\texttt{Names}\xspace}
\newcommand{\body}{\texttt{body}\xspace}
\newcommand{\bodies}{\texttt{bodies}\xspace}
\newcommand{\sep}{\noalign{\vskip 0.8mm}}

%% file: 000abstract.tex
In many domains, a latent competition among different conventions determines which one will come to dominate. 
One sees such effects in the success of community jargon, of competing frames in political rhetoric, or of terminology in technical contexts. These effects have become widespread in the on-line domain, where the ease of information transmission makes them particularly forceful, and where the available data offers the potential to study competition among conventions at a fine-grained level. 

In analyzing the dynamics of conventions over time, however, even with detailed on-line data, one encounters two significant challenges. 
First, as conventions evolve, the underlying substance of
their meaning tends to change as well; and
such substantive changes confound investigations of social effects.
Second, the selection of a convention takes place through the complex
interactions of individuals within a community, and contention
between the users of competing conventions plays a key role in the
convention's evolution.
Any analysis of the overall dynamics must take place in the presence of these two issues. 

In this work we study a setting 
in which we can cleanly
track the competition among conventions while explicitly taking
these sources of complexity into account.
Our analysis is based on the spread of low-level
authoring conventions in the e-print arXiv over 24 years and roughly a
million posted papers: by tracking the spread of macros and other
author-defined conventions, we are able to study conventions that vary
even as the underlying meaning remains constant. We find that the
interaction among co-authors over time plays a crucial role in the
selection of conventions; the distinction between
more and less experienced 
members of the community, and the distinction between conventions with
visible versus invisible effects, are both central to the underlying
processes. Through our analysis we make
predictions at the population level about the ultimate success of different synonymous conventions over time --- and at the individual level about the outcome of ``fights'' between people over convention choices.

%% file: 010intro.tex
% !TEX root = paper-main.tex

\newcommand{\xhdr}[1]{\paragraph*{\bf #1}}
\newcommand{\omt}[1]{}
\def\arxiv{arXiv}

The {\em diffusion of innovations}, a rich area of study with its
roots in sociology, has developed into a lens for addressing the
spread of new ideas, information, behaviors, and technology across
a wide range of domains
\cite{rogers-diffusion}.
In recent years, the on-line domain has provided a powerful setting
in which to study this process. 
The detailed view of the problem offered by such data
has made it possible to provide new
insights into many aspects of diffusion, including its 
temporal properties \cite{crane-youtube},
its structural properties at both a local~\cite{backstrom-kdd06,leskovec-ec06j,ugander-pnas12}
and global level~\cite{adamic-how-met-me,goel-structural-virality,liben-nowell-pnas08},
its level of predictability
\cite{salganik-music,cheng-www14},
and the characteristics of the
participants
\cite{danescu-www13,rotabi-icwsm16}.

One of the central questions that theories of diffusion can address
is the process of competition and selection among conventions:
when there are multiple possible behaviors and a group must choose
among them, can we characterize how this selection process takes place,
and how the latent interaction between competing options unfolds?
The rise of new idioms and terminology
\cite{danescu-www13};
technical standards in engineering and technological domains
\cite{arthur-competing-tech};
themes in political rhetoric \cite{gentzkow-partisan-newspapers}; 
and styles
in artistic and other subjective domains 
\cite{salganik-music}
are all cases where we
can pose such questions. 
It is important to note that the set of issues surrounding 
such conventions are far from monolithic --- in particular, in 
cases with high costs to miscoordination, one tends to see a single
convention crowd out all the others almost completely,
while in cases where the convention has lower coordination effects
and poses lower normative constraints, one typically finds 
extensive coexistence of conventions, with one convention dominating
and others persisting in parts of the population~\cite{young-conventions-jep}.

Our work here begins by noting two central methodological challenges 
that arise in studying the evolution of conventions: one is an issue
of content versus structure, and the other is an issue
of local versus global effects.

$\bullet$ {\em Substantive shift.}
First, it is possible for one
convention to eclipse another because of 
a {\em substantive shift} --- in which the substantive 
meaning
of one convention has a relative advantage over the other.
If we seek to understand the role that social structure plays, we
must look for settings in which the competing conventions are
essentially {\em synonymous} at the level of their substance.
The work on this question to date has faced the challenge that
in most natural settings it is hard to verify whether competing conventions are semantically equivalent, and thus to disentagle social effects from the relative advantage of one convention over another.

This is a distinction that is also familiar in studies of biological evolution,
where the use of {\em neutral variation} --- 
mutations that have minimal
effect on an organisms's fitness --- has come to be an enormously
influential methodology for studying
effects of population structure on evolution in the absence of 
overt selective pressure
\cite{kimura-neutral-theory-book}.
What is the analogue of neutral variation in the diffusion
of on-line information?

$\bullet$ {\em Diffusion through interaction.}
Second, much of the work on diffusion --- both theoretically
and via on-line data --- has studied the global competition among
conventions via local mechanisms of contagion: 
these mechanisms posit that at a local level,
the use of the convention spreads from one person to another,
either probabilistically or through best-response behavior, and the 
competition among these contagion processes leads to the global outcome.
But in most domains where non-trivial conventions are competing, 
the competition takes place not just globally but also locally
through person-to-person interaction.
Returning to our examples above: a single interaction between
speakers in a dialogue or discussion,
collaborators on a technical project, politicians framing a shared
position, or artists performing a shared work may all implicitly
involve competition between the conventions used by the participants
in this interaction.
The global outcome of the competition between two conventions
may emerge from the results of thousands or millions of these
micro-level competitions.
This type of {\em diffusion through interaction} requires a
fine-grained analysis of the local competition, rather than just
a view of the local dynamics as concurrent contagion processes.

\xhdr{The present work}
In this work, we propose an analysis framework for the competition
and selection among conventions that explicitly addresses these issues
of neutral variation and diffusion through interaction.
We do this through a set of novel definitions and measures, together
with a rich source of data that clearly display both notions at work.

Our data comes from a complete snapshot of the source files on the
e-print arXiv, covering 24 years and over a million papers.
We study how low-level authoring conventions emerge through the
collaborations among different overlapping 
sets of co-authors over a multi-year time span.
The source files on the arXiv provide a detailed view into a wide
range of such authoring conventions; we focus primarily on the
role of author-defined macros in \LaTeX\ as one abundant supply of conventions.
In writing the \LaTeX\ source for a paper, authors will often define 
one or more macros as a way to make the writing of the paper easier
and more modular;
as in a standard programming language, 
each instance of a macro instance is specified
by a name (henceforth \name) and a body definition (henceforth \body) 
which defines the functionality of the macro. 
Each time the formatting software for the paper 
sees the name of the macro, it replaces it with the body of the macro 
in the text; thus, for example, the command,
 \texttt{\textbackslash{def\textbackslash{Reals}}\{\textbackslash{mathbb}\{R\}\}} defines a macro, and whenever the author writes 
\texttt{\textbackslash{Reals}}
in the source file for the paper, the symbol
$\mathbb{R}$ will appear in the outcome.

The appeal of focusing on macros is that they provide 
an extremely rich source of synonymous conventions in
the social ecosystem of the arXiv.  Whenever the \name for a macro
changes while the \body remains the same --- for example, when someone
chooses to use \texttt{\textbackslash{R}} 
instead of \texttt{\textbackslash{Reals}}
for the symbol $\mathbb{R}$ --- the author is settling on an arbitrary
choice of convention while the underlying meaning remains constant.
As argued above, this type of control for meaning is crucial if
we want to study the social structure around convention change separately
from substantive shifts in content;
however, controlling for meaning is very hard to achieve unless one
has an almost mechanistic specification of this meaning.
Macros provide us with precisely such a specification 
in their \body.
Also,
they are pervasive in arXiv: roughly 40\% of all arXiv papers contain
at least one user-defined macro.

Moreover, because papers on the arXiv are largely co-authored,
the competition among synonymous macros is also a powerful setting
in which to define and then study some of the basic properties
of diffusion through interaction.
Two macro \names with the same \body are competing not just globally
based on their relative prevalence in the full population of papers,
but also locally each time two people who follow different conventions
come together to co-author a paper.
Analyzing the history of the arXiv provides us with a way to
study how such instance-by-instance competition plays out in the
context of these larger diffusion processes.

The arXiv thus provides us with the ingredients for analyzing 
information diffusion in a way that addresses these methodological 
challenges.  At the same time, we note that the arXiv is of course
a controlled domain representing a single type of broad activity ---
scientific authorship --- and as such our work is approaching these issues
via a case study of this particular domain. 
It will be interesting to study how the observations here generalize
to different contexts; our approach is set up to facilitate this by
providing a road map for these types of analyses across domains.

\xhdr{Overview of results}
We begin by using the 
controlled
 setting provided by our data to
study the competition between synonymous conventions at a global level.
A concrete way to formulate this question is to look at 
two competing \names for the same macro \body up to a certain point
in time, and ask whether we can predict which \name will become
dominant at some point in the future.
First, we find that properties of the \name itself --- e.g., features
related to its orthography, since the meaning is fixed ---
do not seem to have any predictive power; the differences in the
competing \names for the same macro \body appear to truly represent
neutral variation, a fact that offers a striking opportunity to explore
other features in the absence of selective pressure.\footnote{The 
fact that the \name itself provides no predictive power may be in
part a reflection of the fact that authors tend to choose reasonable and
informative \names for their macros; it is easy to imagine that
a particularly inapt \name could have more difficulties in its adoption,
but this is not the situation that generally seems to apply.}

We find, instead, that features related to the {\em experience}
of a \name's early users --- 
the number of previous papers that each has written on the arXiv ---
have significant predictive value for the question of whether
a macro \name will grow to become dominant.
In general, \names that eventually become dominant tend to start with an initial
author population that is relatively ``younger'' (with lower experience),
and then they successfully 
spread to 
``older'' users.
\Names that don't achieve dominance are more associated with
initial user populations that are older in aggregate, 
and also 
fail to spread
to new adopters with
higher experience.
These hand-offs between different 
``generations'' of people, and how they contribute to the success of
a convention, is an interesting issue connected to the role of
status in diffusion \cite{danescu-www13,rogers-diffusion}, and the results
from our data suggest interesting directions in which to explore
these issues further.

We next ask how this competition plays out at a local level,
dropping down from the global scale in order to explore 
diffusion through interaction.
We develop a framework for analyzing the 
instance-by-instance competition that arises when authors following 
different synonymous conventions meet to collaborate.
In particular, if authors $A$ and $B$ meet for the first time to
write a two-authored paper, and they have previously used
different macro \names for the same macro \body, how does the
resulting ``fight'' over the choice of convention turn out?
We find that the relative experience level of the two authors 
is again a highly informative property of the interaction;
the author who is ``younger'' (with lower experience) tends to
win these fights, with the probability of winning increasing as
the gap in experience grows.
Building a set of features based on experience --- both numerically
and through certain more complex structural analogues, such as the
authors' graph-theoretic properties in the larger co-author network ---
we are able to develop 
methods
 for predicting the outcomes of
these fights with non-trivial accuracy.

It is an interesting question to consider possible mechanisms for
the dominance of younger participants in these instance-by-instance 
fights; a natural hypothesis is that they play a larger role in the
detailed implementation of the paper, and hence have more control
over definitional questions such as macros.
Such a model would suggest the conjecture that 
younger co-authors should not necessarily win fights over questions
that are less about low-level implementation and more about 
high-level, visible decisions where the status of the older co-author
is arguably more implicated.
We show that this indeed appears to be the case, by studying the latent
competition between co-authors over conventions in the title of the
paper, rather than the macros.
We can think of the title as occupying the opposite end of the visibility
spectrum from macro names, in that decisions about titling conventions
are highly visible; and here, under a set of definitions that we 
formulate in the paper, we find that the older co-author tends to win
fights about titling conventions, with the effect increasing as the
experience gap increases (here in the opposite direction from
what we saw in fights over macro \names).
In summary, our results suggest the beginnings of a set of principles
that could be summarized in caricature as, 
``In a collaboration, the younger people win
the invisible fights while the older people win the visible fights.''
We argue that developing this notion
more deeply is an interesting direction for
further research, and we point to some additional steps along these lines.

In summary, our focus is on developing new definitions 
around some fundamental issues that have been difficult to address
in diffusion and the selection of conventions ---
the role of neutral variation, represented through the
properties of synonymous conventions, and the dynamics of 
competition not just at a global level but through the 
continuous low-level competition between users of different conventions
as they interact in the system.
We hope that our exploration of these definitions and concepts in
a case study on the arXiv will indicate how such analyses can be
carried more broadly across other domains as well.

%% file: 020related.tex
% !TEX root = paper-main.tex
\xhdr{Words as conventions} One of the most widespread 
sources
 of conventions
is in the choice of words used to refer to particular concepts.
As noted above, our use of macros is designed to contrast with
an inherent source of
complexity in the analysis of conventions in natural language, namely
the severe difficulty in controlling for the precise meaning of a concept
as the words referring to it change.

Analysis of changes in language over long time periods has considered a dual
problem to ours:  how fixed linguistic constructs acquire new meanings. 
This has been undertaken recently 
in studies of historical shifts in word meanings
\cite{tahmasebi2011towards,Hamilton:Acl:2016} and
grammatical constructions \cite{Perek:Acl:2014}, relying on books and
news data that span long periods of time.
Related studies have been performed in the on-line domain, analyzing
global changes in the linguistic system of Twitter
\cite{Eisenstein:ProceedingsOfNaaclHlt:2013,Eisenstein:ProceedingsOfTheNaaclWorkshopOnLanguage:2013,Goel:InternationalConferenceOnSocialInformatics:2016}
and other on-line communities~\cite{danescu-www13,springerlink:10.1007/9780387304243214}.  

Sociolinguistic studies of linguistic change have addressed changes in phonology and spelling that vary systematically across time \cite{Labov:Language:2013}, status \cite{william1966social} and region \cite{Peersman:ArxivPreprintArxiv160102431:2016}.  As such, these studies are similar to ours in that they also explore variations in form of conventions used to refer to fixed concepts (e.g., whether the final `r' in `car' is pronounced or not), albeit the discussions are generally limited to a handful of examples.

\xhdr{Diffusion of information and cultural items}
As discussed in the introduction, there has been a long line of work
studying the processes by which discrete units of information diffuse
on-line; these include memes
\cite{Simmons:ProceedingsOfIcwsm:2011,Omodei:ProceedingsOfAseIeeeSocialcom:2012},
hashtags on Twitter
\cite{Romero:ProceedingsOfThe20ThInternationalConferenceOn:2011,Romero:ProceedingsOfIcwsm:2013,Maity:ProceedingsOfCscw:2016}
and on-line news content
\cite{adamic2005political,Berger:JournalOfMarketingResearch:2012,Bakshy:Science:2015}.
A growing strand of research within this topic has considered 
the problem of predicting future popularity, with specific prediction
studies involving 
downloadable content~\cite{salganik-music},
quotes embedded in broader cultural contexts
\cite{DanescuNiculescuMizil:ProceedingsOfTheAcl:2012,Bendersky:ComputationalLinguisticsForLiteratureWorkshopAtNaacl:2012},
hashtags \cite{Tsur:ProceedingsOfWsdm:2012}, and
memes \cite{cheng-www14,tsur2015don}.

Most of these previous studies could not
control for the meaning of the convention or content that is spreading, 
with two notable exceptions. 
The first is a study on the
emergence of the retweet convention on Twitter
\cite{kooti2012emergence}; this represents an in-depth study of a
single convention providing extended insights, in contrast to 
our study of thousands of distinct conventions and the common properties
across them.
The second is a study of competition between hashtags
on Twitter that only differ in capitalization, suffixes, or 
relative levels of abbreviation (e.g.,
{\em \#saveTheNationalHealthService} vs. {\em \#savethenationalhealthservice}
vs. {\em \#savetheNHS}) \cite{tsur2015don}.
Our setting allows for a more general way of identifying synonymous
conventions, and for verifying that they are indeed synonymous.
And perhaps more crucially, the study of hashtags in \cite{tsur2015don}
showed that the orthography of different hashtags was in fact 
predictive of their success, establishing that in fact these different versions
of hashtags do not represent neutral variations as in our case, but
instead variation that affects relative fitness.

\xhdr{Role of experience}
Our work also explores the interplay between individuals' levels of
experience and their roles in the diffusion of conventions,
including the question of whether new conventions originate with
younger members of the community, or whether the older members
have a relative advantage in imposing their forms of conventions.
Such questions about trade-offs based on experience and
status in the diffusion of innovations has a long history
of study in off-line domains
\cite{deutschmann-adopter-characteristics,rogers-adopter-characteristics,simmel-sociology,mclaughlin-change-agent-revisited,valente-network-interventions,burt-good-ideas,krackhardt-org-viscosity}, and more recently 
has been explored in on-line domains as well 
\cite{danescu-www13,rotabi-icwsm16}.
However, these lines of work do not look at instances where
synonymous conventions compete with each other, or where
it is possible to see such competition playing out at a local level
through person-to-person contention.

%% file: 030data.tex
% !TEX root = paper-main.tex

%

As discussed in the introduction, we use a dataset of macros from
the e-print arXiv, so as to be able to look at variations in 
language conventions (the \names of macros)
while controlling for their meaning (the \body).

The arXiv is a repository of scientific pre-prints (covering physics,
mathematics, computer science, and an expanding set of other
scientific fields). 
The arXiv contains the source files for almost all the papers that have
been uploaded to it, with most of these written in \LaTeX.
Our dataset consists of the full source of all \LaTeX \xspace files on the arXiv,
from its inception (July 1991) through September 2015, a corpus of over
a million papers.
For a prefix of this time period, the ordering of the papers in our data is
only resolved up to one-month granularity (the remainder is totally ordered),
but our methods work with this level of granularity.

From the \LaTeX \xspace source files we extract all macros defined by the most
common methods, specifically \texttt{\textbackslash def},
\texttt{\textbackslash newcommand} and \texttt{\textbackslash renewcommand}.
This results in macros from over $400,000$ papers.
Note that we do not recursively substitute \names that occur inside
of another macro \body.
Table \ref{table:DataBasicInfo} summarizes 
basic statistics
about our data.\footnote{Our macro dataset is available at \url{http://github.com/CornellNLP/Macros}; a repository of arXiv papers is available at \url{http://arxiv.org/help/bulk_data_s3}.}

\begin{table}[ht!]
\centering
\caption{Dataset details}
\begin{tabular}{lr}
\toprule
Number of papers with a macro & 583,078 \\ 
Number of macros defined & 22,628,300 \\ 
Number of unique macro \bodies & 2,586,548\\ 
Average number of \names per \body & 1.40 \\
\sep
Number of unique authors & 222,689 \\ 
Average author per paper & 2.35 \\ 
    \bottomrule

\end{tabular}
\label{table:DataBasicInfo}
\end{table}

%% file: 040method.tex
% !TEX root = paper-main.tex

\input{042changeover}
\input{041fights}

%% file: 042changeover.tex
% !TEX root = paper-main.tex
\section{Global Competition Between\\ Conventions}
\label{sec:changeover}

We begin by considering the competition between conventions, in the form
of macro names, at a global level.
To give a concrete sense for the behavior we are interested in studying,
here is a simple example of competition among macro names --- 
one of many with a similar flavor on the arXiv.
In March 1996, Luty, Schmaltz, and Terning posted a paper to the arXiv,
on an application of gauge theories in theoretical physics, in which they
defined the macro name {\texttt {\textbackslash Yfund}} to expand
to a macro body representing a very simple instance of a combinatorial
structure known as a Young tableau:
\begin{quote}
{\texttt {\textbackslash raisebox\{-.5pt\}\{\textbackslash drawsquare\{6.5\}\{0.4\}\}}}
\end{quote}
This macro body was used again (with the same name 
{\texttt {\textbackslash Yfund}}) in two more papers in May 1996, 
seven more in the remainder of 1996, and a steady stream of others after that.
Of the first 42 uses of this macro body, all but two referred to it
by the name {\texttt {\textbackslash Yfund}}. 
(The other two used {\texttt {\textbackslash fun}}, 
a name that never really caught on.)
But then, in a paper in May 1998, Hanany, Strassler, and Uranga
used the name {\texttt {\textbackslash fund}} to refer to this macro body.
A competition soon broke out between {\texttt {\textbackslash Yfund}}
and {\texttt {\textbackslash fund}}, with {\texttt {\textbackslash fund}} 
gradually becoming more prevalent.
The macro body has by now appeared in over 750 papers on the arXiv;
of the most recent 100 uses, 39 used {\texttt {\textbackslash Yfund}}
and 61 used {\texttt {\textbackslash fund}}.
Figure \ref{fig: yfund} shows how this changeover between the two macro
names took place; the $x$-axis is a time axis,
indexed in order of uses of the macro body, and the $y$-axis
shows a sliding-window average of the fraction of authors using the
names {\texttt {\textbackslash Yfund}} and {\texttt {\textbackslash fund}}
as a function of time.

\begin{figure*}[!th]
            \centering
        \begin{minipage}[c]{0.65\columnwidth}
            \centering
            \includegraphics[width=\linewidth]{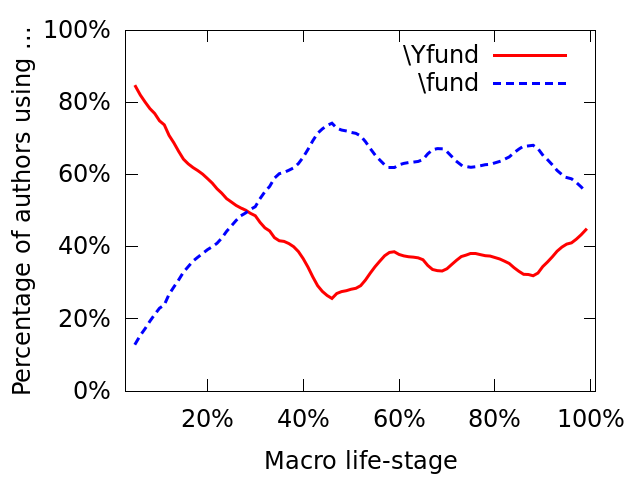}
            \captionsetup{labelformat=empty}
            \caption{(a) Example changeover}\label{fig: yfund}
        \end{minipage}
        % \hfill{}
        \begin{minipage}[c]{0.65\columnwidth}
            \centering
            \includegraphics[width=\linewidth]{{{ConventionChangeOverTime/EarlyLate_SlidingWindow_0.3}}}
            \captionsetup{labelformat=empty}
            \caption{(b) All changeovers aggregated}\label{fig:Change_over_b}
        \end{minipage}
        \begin{minipage}[c]{0.65\columnwidth}
            \centering
            \includegraphics[width=\linewidth]{{{ConventionChangeOverTime/EarlyLate_ChangeOverPoint_0.3}}}
            \captionsetup{labelformat=empty}
            \caption{(c) Crossing-point distribution}\label{fig:Change_over_a}
        \end{minipage}
        \addtocounter{figure}{-1}
        \addtocounter{figure}{-1}
        \addtocounter{figure}{-1}
\caption{Changeover in conventions. 
(a) An example changeover: {\texttt {\textbackslash fund}} surpasses the once-dominant {\texttt {\textbackslash Yfund}} as the preferred \name used to invoke Young tableau; y-axis indicates the percentage of users of each \name out of all authors using the respective \body.
(b) Aggregated temporal usage trends of early name ($N_e$) and late name ($N_\ell$) for all macros undergoing changeovers; their crossing point is well before the middle of their lifespan.
(c) The distribution of crossing points (percentage out of all macros undergoing changeovers).
}
\label{fig: change_over_preliminary}
\end{figure*}

This type of dynamic has played out with many macros on the arXiv,
and the point is not that the choice of macro name is consequential for
the substance of the authors' research.
In fact, the point is the opposite: changes in the macro name are
a source of neutral variation, essentially incidental to the real
progress of the community, and hence they let us probe the changeover
dynamics in conventions
at the level of individuals, their characteristics, and
their interactions.

We also note that the competition underlying the changeover dynamics can
take several different possible forms.  It may be that authors following
the two conventions interact directly through co-authorship or other
mechanisms.  But it may also be that one convention overtakes another
even without direct interaction between the followers of the two conventions,
simply because one of the two conventions grows in adopters and usage
significantly faster than the other.  This too is a form of competition
between the conventions, played out in their relative rates of growth.

\xhdr{Defining changeovers}
We now describe how we identify a broad set of instances in which
one macro \name overtakes another.
For parameters $s$, $q$, and $\theta$, we 
find macro \bodies that have at least $s$ total occurrences,
where there is a \name $N_e$ that is the most used name
in the first $q$ fraction of occurrences, and a different \name
$N_\ell$ is the most used in the last $q$ fraction of occurrences.
Moreover, each of $N_e$ and $N_\ell$ is widely used in the sense
that $N_e$ is used by more than a $\theta$ fraction of authors
who use the macro \body in its first $q$ fraction of occurrences, and
that $N_\ell$ is used by more than a $\theta$ fraction of authors
who use the macro \body in its last $q$ fraction of occurrences.
In our analysis, we use $s = 100$, $q = 0.3$ and $\theta = 0.3$, although
other choices of these parameters produce similar results.

If these properties are met for a given macro \body $\beta$, we say that
$\beta$ undergoes a {\em changeover} from $N_e$ to $N_\ell$,
and we refer to $N_e$ as the {\em \textbf{e}arly name} for $\beta$,
and $N_\ell$ as the {\em \textbf{l}ate name} for $\beta$.
In Figure \ref{fig:Change_over_b} and \ref{fig:Change_over_a} we show some aggregate
properties of the set of changeovers on the arXiv.
First, consider a given macro \body $\beta$ with $m_\beta$ occurrences;
for any $0 \leq t_0 \leq t_1 \leq 1$,
we define the interval $[t_0,t_1]$ in the \body's 
lifespan
to be the set of papers indexed between $t_0 m_\beta$ and $t_1 m_\beta$
in the time-sorted ordering of papers using $\beta$.
We will refer to quantities like $t_0$ and $t_1$ as ``times,'' 
(or ``macro life-stages'')
corresponding to a fraction of the way through a macro \body's lifespan
on the arXiv.
Now, if $\beta$ undergoes a changeover, we define the function
$f_\beta(t,t')$ to be the fraction of authors in the interval
$[t, t']$ that use the early name $N_e$, and 
we define the function
$g_\beta(t,t')$ to be the fraction of authors in the interval
$[t, t']$ that use the late name $N_\ell$.
We can turn these into single-variable functions by fixing an increment
$\delta$ and defining $f_{\beta,\delta}(t) = f_\beta(t,t+\delta)$
and $g_{\beta,\delta}(t) = g_\beta(t,t+\delta)$;
these are just the fractions of usage in length-$\delta$ intervals
beginning at $t$.

Figure \ref{fig:Change_over_b} fixes $\delta = 0.05$
and shows the median values
of $f_{\beta,\delta}(t)$ and $g_{\beta,\delta}(t)$, as functions of $t$,
aggregated over all $\beta$ that undergo changeovers.
It is intuitively sensible that $f_{\beta,\delta}(t)$ should be falling in $t$
and $g_{\beta,\delta}(t)$ should be rising in $t$, since $N_\ell$ is in effect
partially taking over from $N_e$.
It is intriguing, however, that the shapes of the two curves are not
symmetric in 
time --- in that they cross well before the midway point at $t = 0.5$ --- considering that the definition of changeover is temporally symmetrical.

Figure \ref{fig:Change_over_a} shows the distribution
of these crossing points, over all $\beta$ that undergo changeovers.
For this plot, we formalize the crossing point as the minimum $t$
such that $g_{\beta,\delta}(t') \geq f_{\beta,\delta}(t')$ 
for all $t' \in [t, t+.1]$,
so as to require that the crossing persist for a non-trivial interval of time.
This plot too highlights the fact that the crossing tends to occur
early in the usage of the macro \body $\beta$, well before the midway point,
although there is considerable diversity --- for some macro \bodies,
the crossing point comes very late.

\begin{figure*}[!ht]
            \centering
        \begin{minipage}[c]{0.85\columnwidth}
            \centering
            \includegraphics[width=\linewidth]{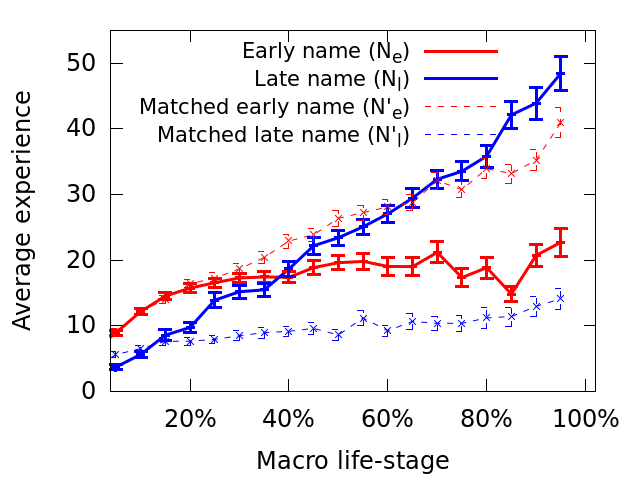}
            \captionsetup{labelformat=empty}
            \caption{(a) Average usage experience}\label{fig:Usage_Adoption_Exp_A}
        \end{minipage}
        % \hfill{}
       \hspace{2.0cm}
        \begin{minipage}[c]{0.85\columnwidth}
            \centering
            \includegraphics[width=\linewidth]{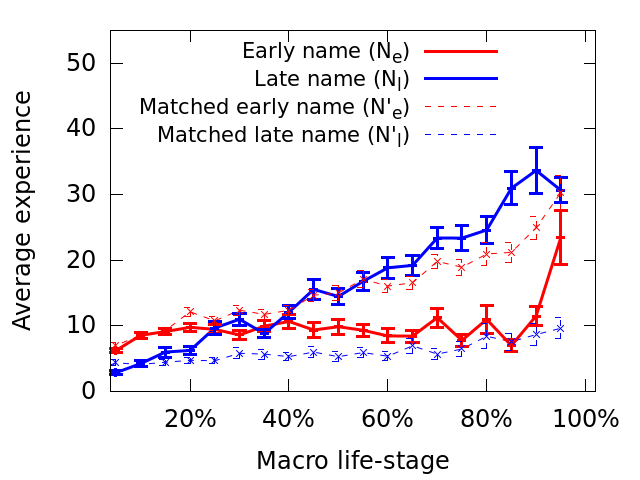}
            \captionsetup{labelformat=empty}
            \caption{(b) Average adoption experience}\label{fig: Usage_Adoption_Exp_B}
        \end{minipage}
        \addtocounter{figure}{-1}
        \addtocounter{figure}{-1}
 \caption{
  Changeovers and user experience. 
 When comparing \names that eventually overtake their competitors ($N_l$, solid blue lines) with those that don't ($N'_l$, dashed blue lines), we observe that they tend to start with a younger user-base and then successfully transition to more experienced users.
 }
    	\label{fig: Usage_Adoption_Exp}
\end{figure*}

\subsection{Properties of the authors in a changeover}
We now 
examine the
 properties of the authors who use competing names
in a changeover.
In order to have a baseline for comparison, 
we pair macros undergoing changeovers with macros that do not undergo changeovers, 
but which
have similar behavior up to their first $q$ fraction of uses.

Thus, for each macro \body $\beta$ that undergoes a changeover,
we find a macro \body $\gamma$ that does {\em not} undergo a changeover,
for which (i) the total volumes of usage
are approximately the same, $m_\beta \approx m_\gamma$, and
(ii) there are two names $N_e'$ and $N_\ell'$ for $\gamma$ such that
the prevalence of these two \names in their early phases are 
approximately the same as $N_e$ and $N_\ell$ respectively:
$f_{\beta}(0,q) \approx f_{\gamma}(0,q)$ and 
$g_{\beta}(0,q) \approx g_{\gamma}(0,q)$.\footnote{The precise
filter we use is to require that $m_\beta/m_\gamma \in [.91,1.1]$
and $|f_{\beta}(0,q) - f_{\gamma}(0,q)|$ and 
$|g_{\beta}(0,q) - g_{\gamma}(0,q)|$ are both below $.01$.}
We will refer to $\beta$ and $\gamma$ as a {\em matched pair}
of macro \bodies.
Intuitively, from the perspective of their volume up to time $q$,
the two \names $N_\ell$ and $N_\ell'$ --- for $\beta$ and $\gamma$ 
respectively --- had very similar initial conditions, and hence
we are forced to look at other properties to find a difference
between them.

A key property that we consider is the {\em experience} of the authors
using these names; recall that an author's experience at a given point
in time is the number of papers they've written up to that time,
which measures a kind of ``age''.
(Accordingly, when we refer to authors as ``younger'' or ``older,''
it is with respect to this measure of experience, and not to
biological age.)
Now, for a matched pair of macro \bodies $\beta$ and $\gamma$, we can
look at the following quantities at a time $t \in [0,1]$:
the average {\em usage experience} of authors of each given name at time $t$,
as well as the average {\em adoption experience} of users of each given
name at time $t$; the former quantity aggregates over 
the experience of all users of
the given name at time $t$, while the latter quantity aggregates 
only over the experience
of authors using the given name {\em for the first time} at $t$.

In Figure \ref{fig: Usage_Adoption_Exp} we show the average usage experience
(left panel) and average adoption experience (right panel) for 
the four names $N_e, N_\ell, N_e'$, and $N_\ell'$, averaged over
all matched pairs $(\beta, \gamma)$.
We notice a few respects in which these curves exhibit similar properties
between $\beta$ and $\gamma$:
first, they all increase, which is natural
since experience values are increasing
as time runs forward on the arXiv.
Moreover, the curves for $N_\ell$ and $N_\ell'$ start out below
the curves for $N_e$ and $N_e'$, which is consistent with 
the intuition that
 new terminology tends to start with more peripheral authors
\cite{rotabi-icwsm16}.

However, the curves for $\beta$ and $\gamma$ also differ in important ways,
and this provides
us with some insight into the differences between macro \bodies 
$\beta$ that undergo changeovers and macro \bodies $\gamma$ that don't.
First, and most visibly, the name $N_\ell$ performs a major transition
over its lifetime, going from authors with very low experience to authors
with very high experience, while the experience of authors using $N_e$
plateaus.  Conversely, $N_\ell'$ fails to perform a corresponding 
transition, and remains concentrated on authors of low experience
throughout its lifespan.
In this sense, $N_\ell$ and $N_e$ almost ``change roles'' as
the plot progresses, with the curve for
$N_\ell$ initially tracking $N_\ell'$ but eventually tracking $N_e'$,
and the opposite holding for $N_e$.
There is also a small but significant difference at the smallest values 
of $t$: the average experience for $N_\ell$ starts out lower than
for $N_\ell'$, a difference that may point to the value of low author
experience in predicting the eventual success of a macro \name.
(We explore this further when we look at interaction dynamics in the
next section.)

\subsection{Predicting changeovers}

We can also evaluate whether the properties we have assembled about the authors
using competing names hold predictive power in the task of forecasting whether a changeover will occur.
We formulate this as a prediction task, where for each matched pair of macro \bodies $\beta$ and $\gamma$ --- each involving two competing \names --- we try to predict early on which of them will undergo a changeover. 
Notice that because of the matching process, we have a balanced dataset
where the two classes have the same aggregate characteristics 
in the early stages of their lifespans.

We first find that using only properties of the two competing macro \names
themselves --- length, number of non-alphabetic characters,
proportions of lowercase and uppercase characters ---
provides no predictive power.
In other words, 
we
 can't predict simply from names like 
{\texttt {\textbackslash Yfund}}
and {\texttt {\textbackslash fund}} which one will prevail.
This reinforces the sense in which these truly represent neutral 
variations; the changes in name do not seem\footnote{There is, however, the possibility that more complex \name features could turn out to hold predictive power.}
 to be fitness-enhancing on
their own.

However, we get non-trivial predictive power when we add 
attributes of the authors using the \names in their early stages.
In particular, we define features based on the number of distinct
authors using each \name in the first $q$ (= 0.3) fraction of the
macro \body's lifespan, as well as the average usage experience and
average adoption experience in windows of 
 $[t,t+.05]$ for
$t \in \{0,.05,.10,.15,.20,.25\}$.

Here and in the rest of the paper, we perform logistic
regression using features that are normalized using the z-score,
with data that has balanced labels and $80$/$20$ split on data 
for training and testing.
 The accuracies using different subsets of
the features are presented in Table \ref{table: change_over_accuracy}.
The most important features are the average usage experience and
the average adoption experience, with low values favoring changeovers:
a \name used by a younger generation is more likely 
to take over its competitor.

\begin{table}[ht!]
\centering
\caption{Accuracy of changeover prediction ($\pm$ 4\% confidence intervals for all rows).}
\begin{tabular}{lc}
\toprule
\textbf{Feature set} & \textbf{Accuracy}
\\ \midrule
Random baseline &  50\%
\\
Name features &  50\%
\\ \sep
Average usage experience features &  58\%
\\ 
Average adoption experience features & 60\%
\\
All author-based features &  59\%
\\ \sep
All features & 57\%
\\ 
\bottomrule
\end{tabular}
\label{table: change_over_accuracy}
\end{table}

%% file: 041fights.tex
% !TEX root = paper-main.tex
\section{Diffusion through Interaction: Dynamics of Local Competition}
\label{sec:macrofights}

We now begin with a set of analyses designed to study how diffusion
through interaction is taking place in our domain --- the way in which
competition over a set of conventions, in the form of different macro
\names for the same macro \body, is taking place at a paper-by-paper level.

The ``age'' of the authors will again play an important role in these analyses,
and we continue to 
use the {\em experience} of an author --- the number of papers they
have written --- as a measure of age.
Unless otherwise specified, when we are considering an author in the context
of a particular paper they have written, we will be thinking about
their experience at the moment this paper was written (as opposed to their
eventual experience at the end of our dataset).

\begin{figure*}[!ht]
        \begin{minipage}[l]{0.65\columnwidth}
            \centering
            \includegraphics[width=\linewidth]{{{Fights/InvisibleFight_ProbOldWinning_VS_ExpDiff}}}
           \captionsetup{labelformat=empty}
            \caption{(a) Invisible fights: \name}\label{fig: fight_A}
        \end{minipage}
        \begin{minipage}[r]{0.65\columnwidth}
            \centering
            \includegraphics[width=\linewidth]{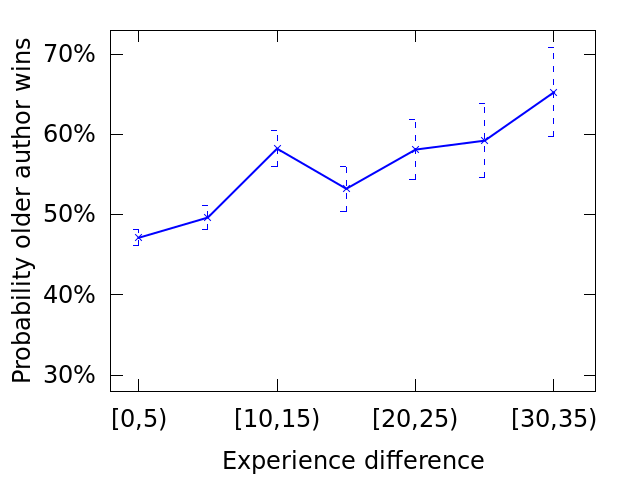}
            \captionsetup{labelformat=empty}
            \caption{(b) Visible fights: paper title}\label{fig: fight_B}
        \end{minipage}
           \begin{minipage}[r]{0.65\columnwidth}
            \centering
            \includegraphics[width=\linewidth]{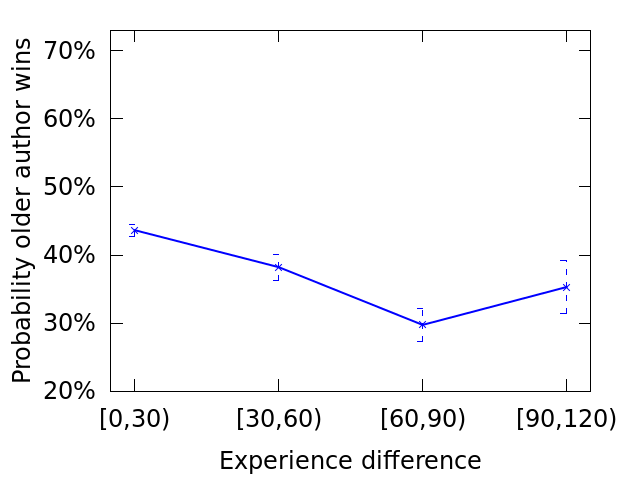}
            \captionsetup{labelformat=empty}
            \caption{(c) Low-visibility fights: \body}\label{fig: fight_c}
        \end{minipage}
        \addtocounter{figure}{-1}
        \addtocounter{figure}{-1}
        \addtocounter{figure}{-1}
 \caption{Percentage of fights won by the older author as a function of difference in  experience.
The larger the experience gap, the more likely the younger author is to win the (invisible) macro \name fights (a) and the (low-visibility) macro \body fights (c); the opposite trend holds for the (much more visible) title fights (b).
 }
 \label{fig: Exp_OldProbability}
\end{figure*}

\subsection{Fights over macro names}

We now consider the outcome of competition between two authors
who have recently used different conventions for the same macro \body.
In order to have a consistent structured
setting for such competition, we consider 
situations in which two authors $A$ and $B$ meet to write a paper, each having used
a different \name for the macro \body $\beta$, and one of these two names is
used in their co-authored paper.
In this case, we say that a {\em fight} has occurred between $A$ and $B$
over the choice of name for \body $\beta$, and the fight is won by the author
whose name is used in their joint paper.
A basic question is to characterize and potentially predict the
winner of a fight of this form --- how does it depend on the properties
of $A$ and $B$, and potentially of the macro \body and \name?

More formally, we require that 
(i) two authors $A$ and $B$
write a paper $P$ in which they
are the only co-authors;\footnote{For robustness, we also redo the analysis on three-author papers, examining fights between the second and the third authors and obtain qualitatively similar results.}
(ii)~the paper $P$ involves a macro \body $\beta$
that each author has used before; (iii) in their most recent 
uses of $\beta$ (in earlier papers), $A$ and $B$ used names
$\eta_A$ and $\eta_B$ with $\eta_A \neq \eta_B$;
and (iv) one of $\eta_A$ or $\eta_B$ is used as the \name for
$\beta$ in the new co-authored paper $P$.
Further, in order to study macros that have non-trivial usage, we require
that the macro \body be used in the dataset by at least 30 distinct authors,
and that the length of the body be at least 10 characters.
Also, since certain pairs of authors might satisfy conditions (i)-(iv)
many times, and there might be close alignment in the outcome of all
their fights, we only consider a single fight for all pairs
of papers that $A$ and $B$ co-author, selecting a
chronologically earliest one.\footnote{For the portion of
our data in which we only have time granularity at the one-month level,
we further restrict to instances in which $A$ and $B$ have 
no other papers in the month of their current co-authorship and
their previous usage of $\beta$, making the temporal ordering
unambiguous.}

The left panel of Figure \ref{fig: Exp_OldProbability}
shows that the younger author (the one with lower experience) wins
these fights significantly more often than the older author,
with the probability that the older wins decreasing 
as the experience difference between the two authors increases.
These results are significant ($p < 0.05$)\footnote{Here and throughout the paper we use the binomial test for statistical significance.} using the full set of
1574 fight instances that passed the conjunction of all the filters 
described above, and was balanced to control for author-position effects, with the first and second authors winning an equal number of fights.

\xhdr{A prediction task}
We can use this distinction in experience to perform a prediction task:
given an instance of a fight, and the past history leading up to it,
how accurately can we predict who will win the fight?

\cut{
The most basic approach to this would be to simply formulate a
prediction problem using two features: 
the experience of the older author and the experience of the younger author.
}
The most basic approach to this would be to simply formulate a
prediction task problem using two features: 
the experience of the first author and the experience of the second author, aiming to guess which of them will win the fights.  
The left panel of Figure \ref{fig: Exp_OldProbability} suggests that we would
already be able to achieve non-trivial performance from just these
two features, and we find that we achieve $58.2\%$ accuracy 
using a logistic regression model.
If we declare an instance to have label 0 when the first-listed
author wins the fight, and label 1 when the second-listed author
wins the fight, then we can interpret the coefficients for the
experience of these two authors in the logistic regression model, shown in
Table \ref{table: invisible_fights_coeffs}.
Note that the positive coefficient for the first author (Experience 1)
means that higher first-author
experience produces an output of the logistic function that
favors  label 1, corresponding to the second author winning the fight.
Conversely, the negative coefficient for the second author (Experience 2) 
means that higher second-author experience favors the label 0,
corresponding to the first author winning.
In consequence, for both authors
higher experience works against them winning the fight,
and lower experience tends to favor them in the prediction.

\begin{table}[ht!]
\centering
\caption{
%
%Coefficients
Feature coefficients
 for predicting macro fights outcome when only using experience.}
\begin{tabular}{ccc}
\toprule
~ & Experience 1 & Experience 2\\ 
Feature coefficient & 0.40 & -0.59\\
\bottomrule
\end{tabular}
\label{table: invisible_fights_coeffs}
\end{table}

\begin{table*}
\centering
\caption{Top 6 feature coefficients for predicting the outcome of macro fights.}
\begin{tabular}{cccccccc}
\toprule
~ & Experience 1 & Experience 2 & Flexibility 1 & Flexibility 2 & Degree 2 & Betweenness 2
\\ 
Feature coefficient & 0.48 &  -0.54 & 0.47 & -0.54 & -0.38 & 0.55
\\ \bottomrule
\end{tabular}
\label{table: invisible_fights_full_coeffs}
\end{table*}

We can achieve non-trivially higher prediction performance by
including a set of other natural features about the instance as follows.
In particular, we use the following set of features.
\begin{itemize}
\setlength\itemsep{0.1em}
\item \textbf{Experience:} 
The number of papers the author
has written prior to the fight.
\item \textbf{Prior uses: } The number of prior papers in which the
author has used the macro \body.
\item \textbf{Flexibility: } The fraction of consecutive uses of the \body
in which the author used two different \names.  (This is a measure
of ``flexibility'' in that it shows the fraction of prior uses of the
\body in which the author changed \names relative to their immediately
preceding use.)
\item \textbf{Degree: } We build a co-author graph on all authors who
have used \body, based only on papers that were written prior to the fight.
This feature is the degree of the author in this co-author graph.
\item \textbf{Betweenness: }The betweenness of the author in the 
co-author graph from the preceding point. (This is one measure of 
how central the author is in the graph.)
\item \textbf{Properties of {\name}: } The length of \name.
\item \textbf{Properties of {\body}: } The length, number of non-alphabetic characters and the maximum depth of curly braces in \body.
\end{itemize}

When we put all these features together, we are able to obtain
67.3\% accuracy via logistic regression.  The coefficients
with the 6 largest absolute values in the model are shown in 
Table \ref{table: invisible_fights_full_coeffs}.
Experience behaves as before; degree is naturally aligned with
experience (since older authors have more time to acquire co-authors);
and flexibility behaves as one would intuitively expect
(since more flexible authors are more likely to also change in the current fight).
It is interesting that high betweenness helps predict that an author
will win a fight, since the results on experience might have 
suggested the opposite intuition.

To summarize, 
as noted in the introduction, it is interesting that the older
author does not play the dominant role --- relative to the younger author ---
in determining the outcome of the convention, given what we know
about the tendency of high-status individuals to drive the outcomes
of interactions \cite{willer-exchange-book}.
A natural hypothesis is that the
older author is ceding control over low-level decisions on conventions
like macro \names to the younger author.
This suggests that we may arguably see a different outcome if we
were to look at fights over decisions that were less low-level,
more visible to readers, and hence more prominent.
In such a case, where the status of the older author is more
implicated by the outcome in the eyes of readers,
is the older author more likely to win the fight?
We now describe an analysis that addresses this contrast.

\subsection{Visible fights}

What would a more visible type of fight look like,
and how does the young-old dynamic work in this case?
We now formalize a set of fights involving one of the most 
visible decisions about a paper --- the choice of title.
For fights involving titles, we need a different set-up than the one
we used for macros, and a more indirect one. 
In the case of macros, the convention was the choice
of a \name for a macro \body, but since paper titles are generally
written in a free-form manner, we need to decide
what the space of possible conventions is.

In order to have a concrete definition to work with, and one that
intuitively has a visible effect on the style of the title, 
we define conventions in the choice of title based on the presence
or absence of certain punctuation, formatting, or parts of speech.
Specifically, for a given title, we ask whether it exhibits
one of seven possible styles (not all mutually exclusive): 
Does it contain a colon, question mark, or mathematical notation;
and is its first word a noun, verb, adjective, or determiner?
For each of these seven questions, 
we say that a title is a {\em positive instance} of the corresponding
style if it contains the indicated feature.

Now, if authors $A$ and $B$ write a paper together,
how do we define the notion of a fight over
one of these stylistic titling conventions?
Asking about the immediately preceding title for each author 
produces data that is too sparse to get meaningful results, and
so instead, we look at each author's lifetime tendency to use
each of the stylistic conventions.

\xhdr{Defining a fight over the title}
Specifically, let us fix one of the stylistic conventions 
$\sigma$ defined above,
and consider a two-authored paper in which the authors have
never written a paper before.\footnote{We also add some additional
filters, including a sufficiently high experience for the older author, and
at least 10 lifetime papers by the younger author that are not written
with the older author.}
Let $E_y$ and $E_o$ denote the
experience of the younger and older authors on this paper respectively,
and let $P_y$ and $P_o$ be the lifetime fraction of papers on
which the younger and older authors used convention $\sigma$,
only considering papers they did not write 
together.\footnote{Since the younger author often has relatively few
papers at the time of the fight, we use the set of all papers written
by each author (not counting their joint papers) to determine these
fractions.  This uses information from the future beyond the paper
itself, but note that we are not using this for a prediction task,
only to determine the relative tendencies of the authors to use
the convention over their lifetimes.}
Finally, we let $I_\sigma$ be an indicator variable equal to $1$ or $0$
depending on the presence or absence of the convention in the 	paper.

Intuitively, we'd like to consider the value of $I_\sigma$ in relation
to which of $P_y$ or $P_o$ is larger.
To have a meaningful baseline for comparison, we match each of our 
fights in pairs: for each fight given by $(E_y,E_o,P_y,P_o,I_\sigma)$,
we find a fight using the same $\sigma$ but a different paper, 
where the values of $P_y$ and $P_o$ are swapped, and where $I_\sigma$
is inverted. That is, we find a fight $(E_y',E_o',P_y',P_o',I_\sigma')$ 
with $P_y' \approx P_o$, and $P_o' \approx P_y$, and 
$I_\sigma' = 1 - I_\sigma$.
Thus, we have a set of matched pairs, where in each pair, one of them
has higher $P_y$, the other has higher $P_o$, and they differ on the
presence or absence of $\sigma$.

What is the effect of this construction?  If in each pair, the instance with 
higher $P_y$ is always the one where $I_\sigma = 1$, it would mean
that $\sigma$ always occurs in the instances where the younger author
has a higher tendency toward $\sigma$; in other words, we'd be able
to perfectly infer the presence or absence of $\sigma$ in each given pair
from the relative values of $P_y$ and $P_o$, with the younger
author playing a greater role driving the presence of $\sigma$.
If in each pair, the instance with higher $P_o$ is always the one
where $I_\sigma = 1$, we would again have perfect prediction
with the older author driving the presence of $\sigma$.
In general, we say that in a single pair, {\em low experience is dominant}
if $I_\sigma = 1$ in the instance with higher $P_y$, and 
{\em high experience is dominant} 
if $I_\sigma = 1$ in the instance with higher $P_o$.
If the $I_\sigma$ values were assigned at random, we'd expect
low experience and high experience to each be dominant in half the pairs.
What do we see in the actual pairs?

We find that in approximately 57\% of the pairs, high experience is
dominant, which at the number of pairs we have is significant 
relative to a random assignment baseline with $p < .001$.
Moreover, we can group the pairs into buckets based on $E_o - E_y$,
and perform this analysis on each bucket separately.
As we see in the middle panel of Figure \ref{fig: Exp_OldProbability},
the extent to which high experience is dominant is increasing
in the experience difference --- the opposite effect from
what we saw in the left panel for fights over macro \names.

Thus, we have a concrete sense in which older authors are winning
visible fights over features of the paper title, 
even though they are losing invisible fights over macro \names.

\subsection{Low visibility fights}

We have now seen the outcomes of two types of fights representing 
opposite extremes of visibility --- fights over macro names, which 
are essentially invisible to readers; and fights over stylistic conventions
in a paper's title, which is extremely visible.
Since younger authors tend to win the invisible fights and lose
the visible fights in these formulations, it becomes natural
to probe the spectrum of possible fights in between these extremes and
thus gain more insight into how the outcome of a fight relates to
its level of visibility.

This is largely an open question, but here we describe one 
initial investigation in this direction.
Consider the case in which two authors meet to write a paper,
and they each use the same macro \name but for different \bodies.
For example, both authors might use \texttt{\textbackslash eps},
but one uses it to mean \texttt{\textbackslash epsilon} ($\epsilon$)
while the other uses it to mean 
\texttt{\textbackslash varepsilon} ($\varepsilon$).
Whose macro \body will end up getting used in the paper they write together?
We will call this a {\em macro-body fight}, and what's interesting
is that it has exactly the structure of our earlier macro-name fights,
except with the roles of the \name and the \body reversed:
now the authors arrive with a shared \name corresponding to different
\bodies, and this contention must be resolved.
An important contrast, however, is that for many macro \bodies,
the outcome of this fight will be visible, albeit often at a very
low level in the formatting and choice of symbols in the paper.
We can therefore think of these as {\em low-visibility fights}, and
can ask whether younger or older authors will tend to win them.

To explore this question, we need to deal with the fact that
not all macro-body fights will have visible effects.
Thus we select a small number of very
common macro names where the effects of different \bodies are
generally visible in the paper.
Specifically, we use the following names:
\texttt{\textbackslash proof}, \texttt{\textbackslash eps} and
\texttt{\textbackslash Re}. 
For these fights we run
the same procedure as for macro-name fights, but we swap the roles
of the \name and \body of the macro, and we
remove the filter of length 20, since \names and \bodies are generally
short in this case.  With these three macro names we end up
with $1092$ fight instances. We observe that the young author wins
$60\%$ of the instances, suggesting that the pattern of outcomes
is closer to what we saw in the invisible macro-name fights.
Grouping the results by the difference in experience, we see 
in the right panel of
Figure \ref{fig: Exp_OldProbability} that these low-visibility
fights follow the same trend as the invisible fights.

%% file: 060discussion.tex
% !TEX root = paper-main.tex

Analyzing the competition among conventions has been
a methodological challenge, because the substance underlying
the convention generally changes, at least to some extent,
together with the convention itself.
We study a setting --- macros on the e-print arXiv --- where it
is possible to fully control for the meaning of the convention
(the \body of the macro) even as the convention itself 
(the choice of \name) is changing.
In the resulting analysis, we focus on two main issues. 
First, we find that instances in which one macro \name convention overtakes
another are characterized by young initial users of the convention
that ultimately succeeds, together with a transitional phase in which
the successful convention spreads to older users.
Second, we consider the local, instance-by-instance competition
among pairs of authors who must resolve contention over the choice
of convention in the process of writing a joint paper.
In this type of {\em diffusion through interaction}, we find
that younger authors tend to win fights (such as for macro \names) 
that do not produce visible consequences, or produce low-visibility
consequences, while older authors tend to win fights (such as for
titling conventions) that produce highly visible consequences.

Having a clean methodology to study conventions that are synonymous
makes possible a number of further directions for research.
First, there are other domains where it should be possible to control
for the meaning of a convention, for example in repositories of source
code where naming conventions can change while the behavior of the code
remains constant.
It is an interesting question to see whether similar phenomena hold
in the dynamics of conventions there.
More generally, it is an interesting question to look for additional
structure beyond the changeover and fight dynamics presented here in the
competition between these conventions.
And finally, it is intriguing to consider using the mathematics
developed around the theory of neutral variation 
\cite{kimura-neutral-theory-book} to begin developing
models for the evolution of synonymous conventions over time.

% other domains - code
% can we identify other structure in the evolution of conventions
% math of neutral variation

%% file: 070acknowledgement.tex
% !TEX root = paper-main.tex
\xhdr{Acknowledgments} 
We are grateful to Paul Ginsparg for his advice and for his help with the arXiv dataset and to Dan Jurafsky and Justine Zhang for helpful discussions. Very few macro fights were lost during this writeup.  
This work was supported in part by ARO, Facebook, Google, the Simons 
Foundation, and a Discovery and Innovation Research Seed Award from
Cornell's OVPR.

% other domains - code
% can we identify other structure in the evolution of conventions
% math of neutral variation